\documentclass[12pt,preprint]{aastex}

\begin{document}

\title{Gamma Ray Astronomy With IceCube}

\author{Francis Halzen and Dan Hooper}

\altaffiltext{1}{Department of Physics, University of Wisconsin, 
1150 University Avenue, Madison, WI  53703; hooper@pheno.physics.wisc.edu;
halzen@pheno.physics.wisc.edu}

\begin{abstract}

We demonstrate that the South Pole kilometer-scale neutrino observatory IceCube can detect multi-TeV gamma rays continuously over a large fraction of the southern sky. While not as sensitive as pointing atmospheric Cerenkov telescopes, IceCube can roughly match the sensitivity of Milagro. Also, IceCube is complementary to Milagro because it will observe, without interruption, a relatively poorly studied fraction of the southern sky. The information which IceCube must record to function as a gamma ray observatory is only the directions and possibly energies of down-going muons.

\end{abstract}


\section{Introduction}

One of the challenges of operating a high-energy neutrino telescope is to reject the background of cosmic ray muons traversing the detector. It is possible to use this background to do TeV-energy gamma ray astronomy. Showers generated by TeV gamma rays from sources in the southern hemisphere can produce muons that penetrate to the depth of the IceCube (Achterberg, $\it{et}\,\it{al.}$ 2003) detector. Although relatively few muons are produced in a gamma ray shower, we will demonstrate that a directional source of TeV photons can produce a measurable excess over the background of muons produced by cosmic ray primaries at a rate of $\sim 2 \rm{KHz}$. By sheer statistics, a Crab-like object, the standard candle of TeV astronomy, will produce a yearly excess of $3$-$5\sigma$, depending on the high-energy characteristics of the source. At TeV energies, sources burst to fluxes as high as ten times the flux of the Crab. We will show that IceCube's sensitivity to TeV gamma rays is comparable to Milagro, although neither can match the instantaneous sensitivity of the new generation of atmospheric Cerenkov telescopes. The latter, however, are only capable of observing a few degree patch of the sky on moonless, clear nights. With a South Pole location, IceCube is unique in that it observes the same sky without interruption, 24 hours per day, 7 days per week. It is most sensitive to sources overhead at the South Pole, a poorly studied portion of the southern sky.

While IceCube is a whole-sky muon and neutrino observatory at energies in excess of $\sim1\,\rm{PeV}$, where the atmospheric background is negligible because of its steep energy spectrum, at TeV energies, the background of down-going cosmic ray muons prevents the detection of neutrinos from Southern sources. To operate as a gamma ray observatory, all IceCube must do is record the directions of the ``background" muons and form a muon sky map of the Southern hemisphere. The muons will be triggered at a rate of close to 2KHz and reconstructed in real time. Eventually other shower properties, such as energy, could be recorded and cuts applied to identify muons of gamma ray origin in the much larger background of cosmic ray induced muons.
 
TeV gamma ray astronomy has developed in several important ways in recent years. In the 1990's, galactic sources, such as the Crab Nebula, and extragalactic sources, such as Markarian 421, were observed to produce TeV photons. Today, 6 sources have been detected by multiple experiments with high significance.  Another eight have been detected by a single experiment with high significance. While sources are few by astronomy standards, their high-energy behavior is fascinating and the accumulating data now defies a comprehensive theoretical explanation (Weekes 2003). After the pioneering work of EGRET (Mukherjee, $\it{et}\,\it{al.}$ 1997 and Sreekumar, $\it{et}\,\it{al.}$ 1997), Whipple (Petry, $\it{et}\,\it{al.}$ 2002) and others, new instrumentation is being developed and deployed such as Milagro (Atkins, $\it{et}\,{al.}$ 2001) and the next generation of atmospheric Cerenkov telescopes (ACTs) including VERITAS (Ong, $\it{et}\,\it{al.}$ 2003), MAGIC (Baixeras, $\it{et}\,{al.}$ 2003), HESS (Hofmann, $\it{et}\,{al.}$ 2002), CANGAROO (Kawachi {\it et al.} 2002 and Okumura {\it et al.} 2002) and others, as well as satellite based detectors such as GLAST (Kamae, $\it{et}\,\it{al.}$ 2003). For a recent review of gamma ray astronomy, see Ong (Ong, 2003).

Presently, only the Milagro experiment is capable of monitoring large portions of the sky for TeV-bright transient sources that may appear, such as gamma ray bursts and blazars. Milagro is, however, only capable of observing approximately a $90^{\circ}$ diameter cone of the sky at one time, leaving many prospective transient sources unobserved in the TeV.

\section{Multi-TeV gamma rays In IceCube}
\label{icecube}

IceCube, now under construction, is designed to observe high-energy muons and showers generated in neutrino interactions inside of or nearby the detector volume. Multi-TeV showers created by photon primaries in the atmosphere will also generate muons, some of which may propagate to the depth of the detector.

The number of muons with energy greater than $E_{\mu}$ produced in a shower with a photon primary of energy $E_{\gamma}$ is well approximated by 
\begin{eqnarray}
N_{\mu}(E_{\gamma}, \ge E_{\mu}) \simeq \frac{2.14 \times 10^{-5}}{\cos \theta} \frac{1}{(E_{\mu}/ \cos \theta)} \frac{E_{\gamma}}{(E_{\mu}/ \cos \theta)}.
\label{muons} 
\end{eqnarray}
$\theta$ is the zenith angle of the shower and all energies are given in units of TeV throughout this paper. This parameterization is based on the results of monte carlo and is accurate for muons with energies $E_{\mu} \le E_{\gamma}/10$ and $E_{\mu} \ge 0.1 \, \rm{TeV}$ (Halzen, Stanev \& Yodh 1996, Halzen, Hikasa \& Stanev 1986).  This parameterization describes muons from the decay of pions produced in the photoproduction of shower photons. Muons from muon pair production or the decay of charm particles, which are not important at the energies of interest, have been neglected. For a complete discussion, see (Drees, Halzen \& Hikasa 1989).

For a muon to be observed, it must reach the detector with energy above the threshold of the experiment. For IceCube, this threshold is $\sim .1 \, \rm{TeV}$. As they travel from the surface to the detector, muons lose energy continuously according to
\begin{equation}
\frac{dE}{dX}=-\alpha - \beta E \ ,
\end{equation}
where $\alpha=2.0 \times 10^{-6}~{\rm TeV}\, {\rm cm}^2/{\rm g}$ and $\beta=4.2 \times
10^{-6}~{\rm cm}^2/{\rm g}$ (Gaisser 1991 and Dutta {\it et al.} 2000). The muon range is then
\begin{equation}  
R_\mu = \frac{1}{\beta} \ln \left[ 
\frac{\alpha + \beta E_\mu}{\alpha + \beta E^{\rm thr}_\mu} \right]
\ .
\label{murange}
\end{equation}
For a vertical and down-going muon to reach a depth of 1400\,m with an energy of 0.1\,TeV, the detector threshold, it must have an energy greater than $\sim 0.56\, \rm{TeV}$ at the surface. This threshold is higher for sources not at the zenith as the required range for such muons is increased to $R_{\mu}/\cos \theta $. Table~\ref{table:one} shows the surface threshold of muons as a function of zenith angle.

The flux of a generic TeV gamma ray emitter can be parameterized as
\begin{equation}
\frac{dN_{\gamma}}{dE_{\gamma}} = \frac{F_{\gamma}}{E^{\alpha}} 10^{-12}\,\rm{cm}^{-2}\rm{s}^{-1}.
\label{spectrum}
\end{equation}
Again, $E_{\gamma}$ is in units of TeV. Typical sources which we consider have values of $\alpha \sim 2$.

To calculate the number of muons observed from a given source, we convolve Eq.~\ref{muons} with Eq.~\ref{spectrum}. The result is the number of muons with energy above the surface threshold from a source
\begin{equation}
N_{\mu}(\ge E_{\mu, \rm{sur}}) \simeq  \int^{E_{\gamma, \rm{max}}}_{E_{\gamma, \rm{min}}} dE_{\gamma} \frac{F_{\gamma} \, 10^{12} }{E^{\alpha}} \, \frac{2.14 \times 10^{-5}}{E_{\mu, \rm{sur}}} \frac{E_{\gamma}}{(E_{\mu, \rm{sur}}/ \cos \theta)},
\label{conv}
\end{equation}
\begin{equation}
\simeq  2 \times 10^{-17} \,\rm{cm}^{-2}\,\rm{s}^{-1}   \, \frac{F_{\gamma}}{\cos \theta} \frac{1}{ (E_{\mu, \rm{sur}}/ \cos \theta)^{\alpha}  } \rm{ln}\bigg(\frac{ E_{\gamma, \rm{max}} }{ E_{\gamma, \rm{min}} }\bigg)\, f,
\end{equation}
where $f$ is a correction factor, approximately given by
\begin{equation}
f = \bigg(\frac{E_{\mu, \rm{sur}}/ \cos \theta}{0.04}\bigg)^{0.53}.
\end{equation}
Eq.~\ref{conv} is only accurate when the integral is performed over many decades of energy. Since this is not necessarily the case for us, $f$ has been introduced to correct for this. This parameterization agrees with explicit monte carlo results (Halzen, Hikasa \& Stanev 1986). We will use $E_{\gamma, \rm{min}} = 10 \times E_{\mu, \rm{sur}}$ neglecting the small probability that lower energy gamma showers produce detected muons.

To determine if the number of muons calculated above is a significant observation, the background from cosmic ray muons must be understood. The background for IceCube is predicted to be approximately given by (Achterberg, $\it{et}\,\it{al.}$ 2003)
\begin{equation}
N_{\rm{bkg}} \simeq 0.5 (\cos \theta)^{2.3} \delta^2 \, t
\end{equation}
where $\delta$ is the angular resolution of the detector. For IceCube, the angular resolution is expected to be better than $1^{\circ}$; we will take $\delta = 1^{\circ}$ as a conservative estimate.

The signal to noise for a given source is then
\begin{equation}
\frac{N_{\rm{sig}}}{\sqrt{N_{\rm{bkg}}}} \simeq 
\frac{1.1 \times 10^{-16}  F_{\gamma} \, (\cos \theta)^{\alpha-1.53} }{E_{\mu, \rm{sur}}^{\alpha -0.53}  } \rm{ln}\bigg(\frac{ E_{\gamma, \rm{max}} }{ E_{\gamma, \rm{min}} } \bigg) A_{\rm{eff}} t \bigg/ \sqrt{0.5 \times 10^7 \, (\cos \theta)^{2.3} \delta^2 \, t},
\end{equation}
\begin{equation}
\simeq 
\frac{1.6 \times 10^{-16}  F_{\gamma} \, (\cos \theta)^{\alpha-2.68} }{E_{\mu, \rm{sur}}^{\alpha -0.53}  } \rm{ln}\bigg(\frac{ E_{\gamma, \rm{max}} }{ E_{\gamma, \rm{min}} } \bigg) A_{\rm{eff}} \sqrt{t} \bigg/  \delta,
\label{ston}
\end{equation}
where $\delta$ is in degrees, $t$ is the observation time in seconds and $A_{\rm{eff}}$ is the effective area of the detector, $10^{10} \, \rm{cm}^2$ in the case of IceCube.

As a first example, we consider a hypothetical TeV-bright object similar to the Crab Nebula, but located at the south pole zenith. With an overhead source, the muon surface threshold is approximately 0.560 TeV. For the Crab, we use the values $F_{\gamma}=50.0$ and $\alpha=2.6$. The signal to noise of such a source in IceCube is 
\begin{equation}
\frac{N_{\rm{sig}}}{\sqrt{N_{\rm{bkg}}}} \sim
1.5 \times \, \rm{ln}\bigg(\frac{ E_{\gamma, \rm{max}} }{5.6 \, \rm{TeV}} \bigg),
\end{equation}
for one year of observation. If the gamma ray spectrum extends to 50 TeV, a 3.3-$\sigma$ observation is predicted. If the spectrum extends to 150 TeV, a 5-$\sigma$ observation is possible. This calculation is somewhat conservative in our choice of $E_{\gamma, \rm{min}}= 10 \, E_{\mu, \rm{sur}}$. Including gamma rays below this energy would enhance the significance of the source.

If a prospective source is not directly overhead, the significance of an observation is reduced. Considering again the Crab-like source, but with $\theta=20^{\circ}$ and $E_{\gamma, \rm{max}}=50 \, \rm{TeV}$, we find a significance of $\sim 2.8$-$\sigma$. For the same source, but with $\theta=45^{\circ}$, a 1.2-$\sigma$ is expected.

There are several sources which are known to be, or are likely to be, TeV emitters in the southern hemisphere including the BL Lac object PKS 2155-304, supernova 1006, the pulsar 1706-44, the radio galaxy Cen-A, the supernova remnant RXJ1713-39, the starburst galaxy NGC 253 and the Vela Pulsar. There are many sources in the southern sky which have not been adequately studied at this time, in part due to the fact that the vast majority of TeV gamma ray experiments have primarily operated in the northern hemisphere. The CANGAROO (Kawachi {\it et al.} 2002 and Okumura {\it et al.} 2002) experiment has established TeV emission from a variety of sources in the southern sky at this time, but like any ACT, cannot possibly monitor large portions of the sky. For a list of hypothetical sources and IceCube's sensitivity, see table~\ref{table:results1}.

In addition to continuous sources, Blazars such as Markarian 421 or Markarian 501 have been observed to flare to luminosities significantly greater than, for example, the Crab. If such an event were to occur in the southern sky, a flare of one week duration and ten times the flux of the Crab, could be observed at $\sim 4.6$-$\sigma$ significance by IceCube.

For particularly bright sources, it may be possible for smaller area neutrino detectors such as AMANDA (Andres, $\it{et}\, \it{al.}$, 2001) or ANTARES (Aslanides, $\it{et}\,\it{al.}$, 1999) to make such observations.  Although these experiments have considerably less effective area, they also have significantly lower energy thresholds. For a summary of high-energy neutrino astronomy, see (Halzen \& Hooper, 2002) or (Learned \& Mannheim, 2000).

\section{Comparison With Milagro}
\label{milagro}

Unlike IceCube, Milagro is located on the surface and, therefore, has a considerably lower energy threshold. However, without the depth of an underground experiment, Milagro suffers from larger backgrounds. 

To establish the sensitivity of Milagro to TeV sources, we consider one of the sources which they have observed with high significance, the Crab Nebula.  The gamma ray spectrum of the Crab is approximately  
\begin{equation}
\frac{dN_{\gamma}}{dE_{\gamma}} \sim 5 \times 10^{-11} E^{-2.6} \, \rm{cm}^{-2} \, \rm{s}^{-1} \, \rm{TeV} 
\end{equation}
which corresponds to $F_{\gamma}\sim 50$ and $\alpha=2.6$ in our parameterization.

We parameterize the effective area of Milagro as (Atkins, $\it{et}\,\,{al}$., 2001)
\begin{equation}
{A_{\rm{eff}}(E_{\gamma})}
=
\left\{\begin{array}{lll}
4 \times 10^{6} \, E^{1.39} \, \rm{cm}^{2}
& \ \ E_{\gamma}>1 \, \rm{TeV},\\
4 \times 10^{6} \, E^{2.35} \, \rm{cm}^{2} & \ \ 1 \, \rm{TeV} > E_{\gamma} > .3 \, \rm{TeV},\\
\sim 0 \, \rm{cm}^{2}
& \ \ .3 \, \rm{TeV} > E_{\gamma}
\end{array}\right. \ .
\end{equation}
The number of signal events predicted from the Crab Nebula in Milagro is then given by
\begin{equation}
N_{\rm{signal}} \simeq \int^{E_{\gamma, \rm{max}}}_{E_{\gamma, \rm{min}}}  A_{\rm{eff}}(E_{\gamma}) \, t \, \frac{dN_{\gamma}}{dE_{\gamma}} \, dE_{\gamma}
\end{equation}
\begin{equation}
\sim 2 \times 10^{-4} \, \times t \,  \bigg[ \int^{E_{\gamma, \rm{max}}}_{1 \, \rm{TeV}} E^{1.39-2.6}_{\gamma} dE_{\gamma} +  \int^{1 \, \rm{TeV}}_{.3 \, \rm{TeV}}  E^{2.35-2.6}_{\gamma} dE_{\gamma}   \bigg] \sim 6 \times 10^{-4} \, \rm{events/sec}
\end{equation}
using $E_{\gamma, \rm{max}}\sim 50 \, \rm{TeV}$. Over one year of observation, with the Crab observable in the sky approximately $20\%$ of the time, we predict $\sim4000$ events/yr in Milagro.   This appears to be a slightly optimistic estimate, as Milagro has actually observed an excess from the Crab of approximately 3870 events/yr. 

The background rate observed in Milagro is approximately 2300 events/day for an object observable $20\%$ of the time.  After 1 year of observation with Milagro, we would expect a signal to noise from the Crab of
\begin{equation}
\frac{N_{\rm{sig}}}{\sqrt{N_{\rm{bkg}}}} \sim \frac{4000}{\sqrt{2300 \times 365}} \simeq 4.4 \, \sigma.
\end{equation}
Using the rate actually observed by Milagro, 3870 events/yr for $\sim900$ days, we get
 \begin{equation}
\frac{N_{\rm{sig}}}{\sqrt{N_{\rm{bkg}}}} \sim \frac{3870 \times 900/365}{\sqrt{2300 \times 900}} \simeq 6.6 \, \sigma,
\end{equation}
which is similar to the 6.4-$\sigma$ result published by the Milagro collaboration (Atkins, $\it{et}\,\,{al}$., 2003 and Atkins, $\it{et}\,\,{al}$., 2001).

To emphasize the comparison with IceCube, for an overhead source with $\alpha\sim2.5$, Milagro is sensitive to objects with $\sim 60\%$ of the minimum flux required by IceCube. As the zenith angle increases, this factor becomes smaller, reaching $\sim20\%$ by $\theta=45^{\circ}$. For a list of hypothetical sources and sensitivities for IceCube and Milagro, see table~\ref{table:results1}. 


\section{Comparison With Atmospheric Cherenkov Telescopes (ACTs)}

Atmospheric Cherenkov Telescopes (ACTs) are designed to be highly sensitive to showers generated by TeV gamma rays. They must be pointed a source, however, and are not capable of continuously observing large portions of the sky as experiments such as IceCube or Milagro can.

For comparison, consider a generic ACT with a threshold energy of .25 TeV and effective area of $3.5 \times 10^8 \, \rm{cm}^2$. The number of events observed for such a telescope is given by
\begin{equation}
N_{\rm{sig}} \simeq A_{\rm{eff}}\, t \, (0.68)^2 \int^{E_{\gamma, \rm{max}}}_{E_{\rm{th}, \gamma}} \frac{dE_{\gamma}}{E^{\alpha}_{\gamma}} \, F_{\gamma}  \, 10^{-12}\, \rm{cm}^{-2} \, \rm{s}^{-1}
\end{equation}
\begin{equation}
\simeq \frac{A_{\rm{eff}}\, t\,  \, F_{\gamma}  \, 4.6 \times 10^{-13}}{\alpha -1 \, E^{\alpha -1}_{\rm{th},\gamma}}
\end{equation}
\begin{equation}
\simeq 6.4 \times 10^{-4} \frac{ F_{\gamma} \, t  }{ (\alpha-1) \times 0.25^{\alpha -1}},  
\end{equation}
where $t$ is the observation time in seconds. The factor of $(0.68)^2$ appears due to the definition of angular acceptance in the estimation of the experiment's effective area.

If the ACT has, for example, a $30\%$ energy resolution and 0.001 steradians angular acceptance, the corresponding background consists largely of misidentified hadronic showers and electron-primary showers. This background can be approximated by
\begin{equation}
N_{\rm{bkg}} \simeq 3 \times 10^{-11} \bigg[ \frac{1}{E_{\gamma, \rm{th}}^{1.7}} + \frac{0.08}{E_{\gamma, \rm{th}}^{2.3}} \bigg] \, A_{\rm{eff}}\, t
\end{equation}
\begin{equation}
\rightarrow  0.13 \,\, \rm{events}\,\,\rm{per} \,\,\rm{second}.
\end{equation}
Combining these results, we get a signal to noise of
\begin{equation}
\frac{N_{\rm{sig}}}{\sqrt{N_{\rm{bkg}}}} \sim
0.0018 \times \frac{ F_{\gamma}}{(\alpha-1) \times 0.25^{\alpha -1}} \times \sqrt{t}.  
\end{equation}
To compare with an experiment such as IceCube and Milagro, consider again a hypothetical, Crab-like TeV-bright source with $F_{\gamma} \sim 50$ and $\alpha \sim 2.6$. In less than 100 seconds of observations time, a 5-$\sigma$ observation could be made.  Similarly impressive results are expected for a variety of sources.

An experiment such as IceCube cannot possibly compete with ACTs in sensitivity to known continuous sources. Unlike ACTs, however, experiments such as IceCube or Milagro are capable of observing large portions of the sky and, therefore, are sensitive to transient sources such as gamma ray bursts or blazars. We will discuss such sources in the following section.

\section{Transient Sources}
\label{results}

In the previous section, we demonstrated that ACTs are far more sensitive to known continuous sources than either IceCube or Milagro In this section, we will discuss the prospects for IceCube and Milagro to observe multi-TeV transient sources to which ACTs are not sensitive.

Using the methods described in sections~\ref{icecube} and~\ref{milagro}, we have calculated the signal to noise expected in both experiments for a variety of hypothetical transient sources. These results (see table~\ref{table:results}) demonstrate Milagro's sensitivity to gamma ray bursts and blazars for values of $F_{\gamma}$ of order $10^4$ and possibly lower, depending on the spectral characteristics and duration of the burst. IceCube, while not as sensitive as Milagro in this respect, is capable of monitoring much of the, unobserved, southern sky for particularly TeV-bright transients. A short duration event ($t\sim\,$ a few seconds), such as a gamma ray burst could be observed with 5-$\sigma$ confidence by IceCube for $F_{\gamma}$ as low as $10^5$ if its spectrum extends to very high energies and is located overhead. For a less ideal short duration source, only extending to tens of TeV, or at $\theta \sim 30^{\circ}-40^{\circ}$, for example, could be observed only if $F_{\gamma}\sim 10^6$, or so. Longer duration transients, such as blazars, which typically have flares of many minutes duration, can be observed with less fluence.

Again, we would like to point out that no experiment other than IceCube will be capable of monitoring the southern sky for TeV-bright transients.

\section{Conclusions}

We have calculated the rates and sensitivities of IceCube as a TeV gamma ray observatory.  IceCube, capable of detecting muons of $\sim .1 \, \rm{TeV}$ and above, can observe the presence of muons generated in multi-TeV gamma ray showers, and distinguish these events from background given a sufficiently TeV-bright source.

We have compared the sensitivities of IceCube, in this respect, to both Milagro and a generic ACT. ACTs are considerably more sensitive, but do not have the ability to monitor large portions of the sky continuously. Comparisons of IceCube with Milagro indicate that IceCube's sensitivity is comparable to Milagro's for sources near the southern zenith. Additionally, these experiments are complementary, as they do not observe the same portions of the sky. If a TeV-bright transient source occurs in the southern hemisphere, IceCube may be the only experiment capable of monitoring it. No additional hardware or software is needed in IceCube beyond its planned design.

\acknowledgements

We would like to thank Ty DeYoung and Gary Hill for valuable discussions. 
This research was supported  by the U.S.~Department of Energy
under grant DE-FG02-95ER40896
and by the Wisconsin Alumni Research Foundation.

\vspace{0.0cm}
 \begin{table}
 \caption{The energy threshold for muons at the surface to reach a detector buried 1400 meters below the Earth's surface in ice, as is the case for IceCube. Sources near the horizon have considerably higher thresholds for muons to successfully propagate to the detector.
 \label{table:one}
 }
 \hspace{4.0cm}
 \begin{tabular} {c c c c c} 
 \hline
&& Zenith Angle
 &$E_{\mu , \rm{sur}} \, \rm{(TeV)}$&\\
 \hline \hline
&&$ 0^{\circ}$&$0.561$&\\
&&$10^{\circ}$&$0.570$&\\
&&$20^{\circ}$&$0.601$&\\
&&$30^{\circ}$&$0.660$&\\
&&$40^{\circ}$&$0.765$&\\
&&$50^{\circ}$&$0.962$&\\
&&$60^{\circ}$&$1.391$&\\
&&$70^{\circ}$&$2.738$&\\
&&$80^{\circ}$&$16.546$&\\
 \hline \hline
 \end{tabular}
 \end{table}


\vspace{0.0cm}
 \begin{table}
 \caption{The significance at which various sources could be observed by IceCube and MILAGRO in one year. All rates are calculated for sources with an $\alpha \simeq 2.6$ spectrum. For MILAGRO, we assumed that each source could be viewed $\sim 20\%$ of the time.
 \label{table:results1}
 }
 \hspace{1.5cm}
 \begin{tabular} {c c c c} 
 \hline
 Source Characteristics
 & IceCube & MILAGRO & \\
 \hline \hline
$F_{\gamma}=1000, \theta=0^{\circ}, E_{\gamma, \rm{max}}=20 \,\,\,\,  \, \rm{TeV}$& $38$-$\sigma$ & $76$-$\sigma$ & \\
$F_{\gamma}=1000, \theta=0^{\circ}, E_{\gamma, \rm{max}}=1000 \, \rm{TeV}$& $160$-$\sigma$ & $110$-$\sigma$ & \\
$F_{\gamma}=1000, \theta=45^{\circ}, E_{\gamma, \rm{max}}=20 \,\,\,\,\, \rm{TeV}$& $12$-$\sigma$ & $76$-$\sigma$ & \\
$F_{\gamma}=1000, \theta=45^{\circ}, E_{\gamma, \rm{max}}=1000 \, \rm{TeV}$& $62$-$\sigma$ & $110$-$\sigma$ & \\
\hline
$F_{\gamma}=100, \theta=0^{\circ}, E_{\gamma, \rm{max}}=20 \,\,\,\,\, \rm{TeV}$& $3.8$-$\sigma$ & $7.6$-$\sigma$ & \\
$F_{\gamma}=100, \theta=0^{\circ}, E_{\gamma, \rm{max}}=1000 \, \rm{TeV}$& $16$-$\sigma$ & $11$-$\sigma$ & \\
$F_{\gamma}=100, \theta=45^{\circ}, E_{\gamma, \rm{max}}=20 \,\,\,\,\, \rm{TeV}$& $1.2$-$\sigma$ & $7.6$-$\sigma$ & \\
$F_{\gamma}=100, \theta=45^{\circ}, E_{\gamma, \rm{max}}=1000 \, \rm{TeV}$& $6.2$-$\sigma$ & $11$-$\sigma$ & \\
\hline
$F_{\gamma}=10, \theta=0^{\circ}, E_{\gamma, \rm{max}}=20 \,\,\,\,\, \rm{TeV}$& $0.38$-$\sigma$ & $0.76$-$\sigma$ & \\
$F_{\gamma}=10, \theta=0^{\circ}, E_{\gamma, \rm{max}}=1000 \, \rm{TeV}$& $1.6$-$\sigma$ & $1.1$-$\sigma$ & \\
$F_{\gamma}=10, \theta=45^{\circ}, E_{\gamma, \rm{max}}=20 \,\,\,\,\, \rm{TeV}$& $0.12$-$\sigma$ & $0.76$-$\sigma$ & \\
$F_{\gamma}=10, \theta=45^{\circ}, E_{\gamma, \rm{max}}=1000 \, \rm{TeV}$& $0.62$-$\sigma$ & $1.1$-$\sigma$ & \\
 \hline \hline
 \end{tabular}
 \end{table}


\vspace{0.0cm}
 \begin{table}
 \caption{The observation time required to make a 5-$\sigma$ observation in IceCube and MILAGRO.  All rates are calculated for sources with an $\alpha \simeq 2.6$ spectrum. 
 \label{table:results}
 }
 \hspace{1.5cm}
 \begin{tabular} {c c c c} 
 \hline
 Source Characteristics
 & IceCube & MILAGRO & \\
 \hline \hline
$F_{\gamma}=10^6, \theta=0^{\circ}, E_{\gamma, \rm{max}}=20 \,\,\,\,  \, \rm{TeV}$& $0.55 \, \rm{sec}$ & $0.027 \, \rm{sec}$ & \\
$F_{\gamma}=10^6, \theta=0^{\circ}, E_{\gamma, \rm{max}}=1000 \, \rm{TeV}$& $0.031 \, \rm{sec}$ & $0.012 \, \rm{sec}$ & \\
$F_{\gamma}=10^6, \theta=45^{\circ}, E_{\gamma, \rm{max}}=20 \,\,\,\,\, \rm{TeV}$& $5.5 \, \rm{sec}$ & $0.027 \, \rm{sec}$ & \\
$F_{\gamma}=10^6, \theta=45^{\circ}, E_{\gamma, \rm{max}}=1000 \, \rm{TeV}$& $0.21 \, \rm{sec}$ & $0.012 \, \rm{sec}$ & \\
\hline
$F_{\gamma}=10^5, \theta=0^{\circ}, E_{\gamma, \rm{max}}=20 \,\,\,\,\, \rm{TeV}$& $55 \, \rm{sec}$ & $2.7 \, \rm{sec}$ & \\
$F_{\gamma}=10^5, \theta=0^{\circ}, E_{\gamma, \rm{max}}=1000 \, \rm{TeV}$& $3.1 \, \rm{sec}$ & $1.2 \, \rm{sec}$ & \\
$F_{\gamma}=10^5, \theta=45^{\circ}, E_{\gamma, \rm{max}}=20 \,\,\,\,\, \rm{TeV}$& $550 \, \rm{sec}$ & $2.7 \, \rm{sec}$ & \\
$F_{\gamma}=10^5, \theta=45^{\circ}, E_{\gamma, \rm{max}}=1000 \, \rm{TeV}$& $21 \, \rm{sec}$ & $1.2 \, \rm{sec}$ & \\
\hline
$F_{\gamma}=10^4, \theta=0^{\circ}, E_{\gamma, \rm{max}}=20 \,\,\,\,\, \rm{TeV}$& $5500 \, \rm{sec}$ & $270 \, \rm{sec}$ & \\
$F_{\gamma}=10^4, \theta=0^{\circ}, E_{\gamma, \rm{max}}=1000 \, \rm{TeV}$& $310 \, \rm{sec}$ & $120 \, \rm{sec}$ & \\
$F_{\gamma}=10^4, \theta=45^{\circ}, E_{\gamma, \rm{max}}=20 \,\,\,\,\, \rm{TeV}$& $55,000 \, \rm{sec}$ & $270 \, \rm{sec}$ & \\
$F_{\gamma}=10^4, \theta=45^{\circ}, E_{\gamma, \rm{max}}=1000 \, \rm{TeV}$& $2100 \, \rm{sec}$ & $120 \, \rm{sec}$ & \\
 \hline \hline
 \end{tabular}
 \end{table}

\end{document}